\documentclass[11pt]{article}%

\usepackage{amssymb}
\usepackage{amsmath}
\usepackage{graphics}
\usepackage{epsfig}
\usepackage{amsfonts}
\usepackage{graphicx}%

\usepackage{subfig}

\usepackage{booktabs}

\newcommand{\eqb}{\begin{equation}}
\newcommand{\eqe}{\end{equation}}
\newcommand{\dmb}{\begin{displaymath}}
\newcommand{\dme}{\end{displaymath}}

\newcommand{\eab}{\begin{eqnarray}}
\newcommand{\eae}{\end{eqnarray}}

\newcommand{\be}{\begin{equation}}
\newcommand{\ee}{\end{equation}}

\begin{document}
\begin{titlepage}
\begin{flushright}
\end{flushright}
\vspace{0.6cm}
\begin{center}
\huge{Finite- and infinite-volume thermodynamics around the zero of the pressure in 
deconfining SU(2) Quantum Yang-Mills theory} 
\end{center}
\vspace{0.5cm}
\begin{center}
\large{Ralf Hofmann}
\end{center}
\vspace{0.1cm}
\begin{center}
{\em Institut f\"ur Theoretische Physik, Universit\"at Heidelberg,\\ 
Philosophenweg 16, D-69120 Heidelberg, Germany;\\ r.hofmann@thphys.uni-heidelberg.de
}
\end{center}
\vspace{1.5cm}
\begin{abstract}
We re-address the self-intersection region in a figure-eight shaped center-vortex loop containing a frequently perturbed {\sl BPS monopole} subject to a core-oscillation 
frequency $\omega_0$, rectifying a numerical error in estimating the region's radius $r_0$ in comparison to the spatial coarse-graining scale of infinite-volume thermodynamics. Implications are discussed. We also compute the lowest frequency $\Omega_0$ of a spherically symmetric plasma oscillation within a {\sl neutral} and spatially homogeneous ball-like region of deconfining phase in dependence of its radius $R_0$. For 
$r_0=R_0$ we compare $\omega_0$ with $\Omega_0$. We point out how 
the idealisations, which are assumed in this work, will have to 
be relaxed in order to address the emission of electromagnetic radiation 
and of non-intersecting as well as self-intersecting center-vortex 
loops away from the surface region of macroscopically sized plasma balls.
\end{abstract}
  
\end{titlepage}

\section{Introduction}

The phase structure and the deconfining thermal ground state of (electric-magnetic dually interpreted) 
SU(2) Yang-Mills thermodynamics suggests the existence of a finite-extent particle with a 
magnetic moment provided by the electric flux along a figure-eight-like 
configuration of a self-intersecting center-vortex loop, 
immersed into the confining phase. This soliton is stable, 
possesses a mass $m_0$ mainly arising from the deconfining energy density of the ball-like, 
Bohr-radius sized self-intersection region whose radius $r_0$ is about fifty times larger than the Compton 
wave length $l_C$. This object exhibits one unit of 
electric charge, immersed into this deconfining region as a {\sl BPS monopole} of mass $m_m$ whose core localises  
to within a radius matching $l_C$, which frequently and indeterministically is relocated and 
forced to oscillate at frequency $\omega_0$ \cite{FodorRacz2004,Forgacs} 
by the action of a trivial-holonomy caloron/anticaloron center \cite{HS1977} in the deconfining thermal ground state \cite{Hofmann2017}. Such a model of the electron (and other charged leptons) 
concretely represents Louis de Broglie's ideas on 
the (quantum) thermodynamics of the isolated particle \cite{deBroglieDoc,DeBroglieRev}, based on the important observation that its 
mass $m_0$ must be related to a time-periodic 
phenomenon of (circular) frequency $\omega_0$ as $m_0 c^2=
\hbar\omega_0$ -- $c$ the speed of light in vacuum, $\hbar$ Planck's reduced unit of action -- 
in the rest frame of this particle. Note that a Lorentz 
boost generates out of a spatially flat oscillation a propagating 
wave, whose (de Broglie) wavelength $\lambda$ 
relates to the spatial momentum $p$ of the particle as $\lambda=\frac{2\pi \hbar}{p}$, and the 
increase of rest energy $m_0 c^2$ by virtue of the boost can be decomposed into a reduction of 
internal heat plus the mechanical work invested. 
The reason why this model of the electron complies with the particle's apparent structurelessness of 
charge distribution, as inferred from scattering experiments with energy-momentum 
transfers up to several TeV, is the 
thermal nature of the self-intersection region (maximum entropy), the locus of 
electric charge essentially being undetermined within the large 
volume $\sim \frac{4}{3}\pi r_0^3$. On the scale of the electron mass $m_0$ 
the center-vortex loop without a region of self-intersection 
is massless and prone to curve shrinking \cite{MoosmannHofmannI}. Emergent in the confining 
phase of the same SU(2) Yang-Mills theory, such a soliton is a natural candidate for the (electron) 
neutrino.     

There are implications of this model for high-temperature plasma physics 
(in considering transport properties in view of the turbulent behaviour induced by unstable 
center-vortex loops  with higher self-intersection numbers \cite{LowTYM,BenderWu}) and for the physics of strongly correlated electrons in a spatial plane (parts of the correlations intrinsically arising from the non-local magnetic 
interactions of vortex sectors \cite{MoosmannHofmannI} and the thermal, indeterministic nature of 
the electric charge density \cite{Hofmann2017}) underlying, e.g., the phenomenon of high-$T_c$ superconductivity. Both fields of research require long-term efforts, however. 

One concern of the present paper is to discuss a few amendments to \cite{Hofmann2017} where the above-sketched model of the electron was discussed. In particular, we 
re-state results on $r_0$ and $m_m$ and the ratio $m_m/m_0$ under the use of the 
proper (and not erroneous as in \cite{Hofmann2017}) monopole mass 
formula $m_m=\frac{4\pi}{e(T_0)} H_\infty(T_0)$ with 
$e(T_0)=12.96$ the value of the effective gauge coupling at the 
temperature $T_0=1.32\,T_c$ ($T_c$ the critical temperature for the deconfining-preconfining phase transition), where the deconfining pressure $P$ vanishes. Here $H_\infty(T_0)=\pi T_0$ denotes the 
asymptotic modulus of the adjoint Higgs field in the BPS 
monopole configuration, linking it to (anti)caloron 
dissociation at maximum non-trivial holonomy 
\cite{Nahm,KraanVanBaal,LeeLu,Diakonov2004,HofmannBook} (and therefore electron-positron pair creation).

There is no essential change in numerical 
values up to this point. However, in computing the ratio $\frac{r_0}{|\phi|^{-1}(T_0)}$ a substantial 
error was committed in \cite{Hofmann2017}: Instead of $\frac{r_0}{|\phi|^{-1}(T_0)}\sim 160$ the actual 
value turns out to be $\frac{r_0}{|\phi|^{-1}(T_0)}\sim 0.104$ which corresponds to 
$\frac{r_0}{T_0^{-1}}\sim 1.29$. (The radius $r_0$ is denoted as $R_0$ in \cite{Hofmann2017}.) 
Therefore, it is incorrect 
to state, as was done in \cite{Hofmann2017}, 
that the self-intersection region of radius $r_0$ represents infinite-volume thermodynamics. 
Rather, this region is deeply contained within the center of the {\sl accomodating} 
caloron or anticaloron, and, judged by the approximate attainment of the asymptotic harmonic dependence of 
$\phi$ on Euclidean time when integrating the according field-strength correlator on a caloron or anticaloron in singular gauge 
over the instanton-scale parameter \cite{HofmannBook}, the use of infinite-volume 
thermodynamics can thus be considered a coarse resemblence only.   

The other objective of our present work is a discussion of an extended, neutral plasma 
ball at $T_0$, immersed into the confining phase, where infinite-volume 
thermodynamics is well saturated in the bulk (disregarding the non-thermal situation 
within a thin shell of preconfining/turbulent confining phase leading to surface-tension 
effects) for a sufficiently large ball radius $R_0$. In particular, 
we ask what the lowest frequency $\Omega_0$ is, associated with a spherically 
symmetric breathing mode describing a homogeneous plasma oscillation about $T_0$. 
This requires the determination of the longitudinal sound speed $c_s$ at 
$T_0$ which turns out to be $c_s=0.479\,c$. Note that, in spite of the vanishing pressure 
at $T_0$, exhibiting a cancellation of the contributions from the thermal ground state and its (partially massive) excitations, this is not much below the ultrarelativistic-gas limit $c_s=\frac{1}{\sqrt{3}}\,c\sim 0.577\,c$.                Subsequently, we then compare $\Omega_0$ and $\omega_0=m_0c^2/\hbar$ at $R_0=r_0$. 

This paper is organised as follows. In Sec.\,\ref{RE} we re-address the computation of $r_0$ using the proper expression for the monopole mass $m_m$. The value of $r_0$ is not affected much compared to the one obtained in \cite{Hofmann2017}. We also compute the ratio of $m_m$ to $m_0$. However, due to a calculational error the ratio of $r_0$ to the coarse-graining scale $|\phi|^{-1}$ turns to out be much smaller 
than what was found in \cite{Hofmann2017} and what was taken as a proof that infinite-volume thermodynamics 
is valid at face value. We argue, however, that the ratio $r_0/T_0^{-1}$ permits an approximately 
thermodynamical treatment because the asymptotic harmonic time dependence of the integral over the field-strength correlator, rendering the Euclidean time dependence of field $\phi$ a mere gauge 
choice \cite{HofmannBook}, is reasonably well approached when cutting off the instanton-scale-parameter 
integral at $r_0$. In Sec.\,\ref{BreathingMode} we address a situation where the deconfining, homogeneous 
plasma, considered at temperatures around $T_0$, is void of an explicit monopole but 
undergoes spherically symmetric, oscillatory breathing by virtue 
of a finite longitudinal sound velocity $c_s$. We reliably compute 
the square of $c_s$ by appealing to the derivatives w.r.t. temperature of the pressure and the energy density, employing the evolution equation for the mass of the off-Cartan modes. 
Both situations, monopole driven oscillation 
and spherically symmetric breathing, are compared in terms 
of their frequencies, the former exhibiting a much more 
rapid oscillation than the latter. Finally, in Sec.\,\ref{Disc} 
we summarise and discuss our results and provide an outlook on future work.

\section{Self-intersection region of a figure-eight shaped center-vortex loop\label{RE}}

In \cite{Hofmann2017} we have proposed a model of the free electron, based on the phase structure 
of SU(2) Yang-Mills thermodynamics and the work in \cite{FodorRacz2004,Forgacs}, 
investigating the response of a BPS monopole to a spherically symmetric initial perturbation 
and the spectrum of normal modes. Here we would like to correct some numerical 
statements arising from an incorrect monopole mass formula (Eq.\,(18) of \cite{Hofmann2017}). Also, 
we point out an error in Eq.\,(21) of \cite{Hofmann2017}, entailing a conceptual re-interpretation 
of the physics associated with the self-intersection region. From now on we work in super-natural units: $c=\hbar=k_B=1$ where $k_B$ denotes Boltzmann's constant.    

\begin{figure}
\centering
\includegraphics[width=15cm]{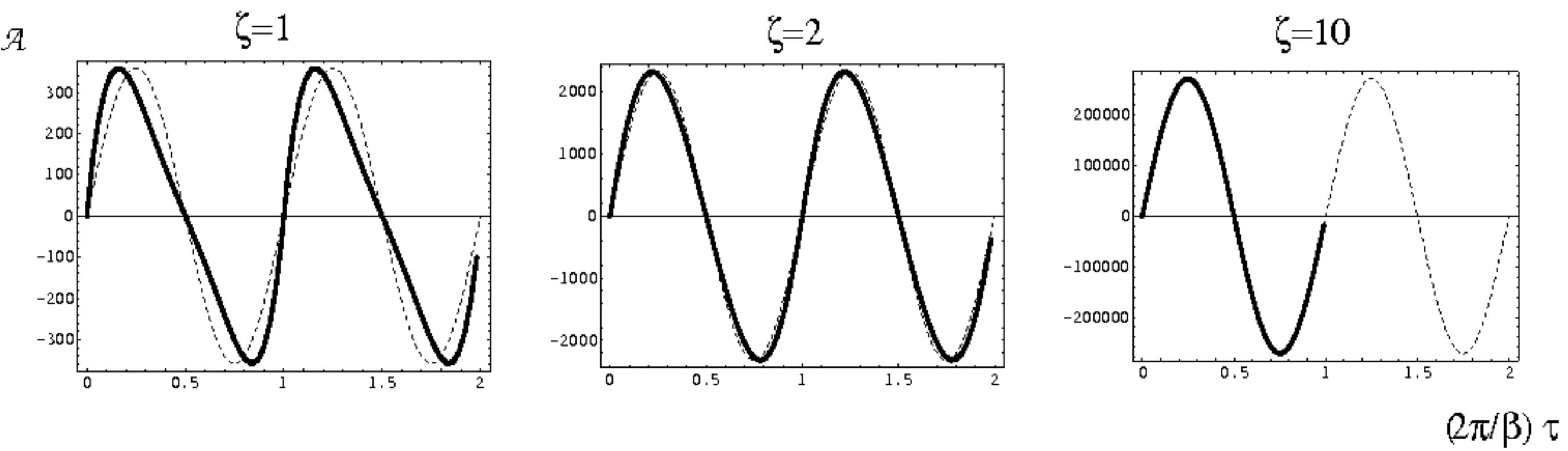}
\caption{Saturation towards a harmonic Euclidean time dependence of the contribution of a Harrington-Shepard 
caloron to the field-strength correlation defining the phase of the field $\phi$ as a function of the scaled cutoff $\xi\equiv \frac{\rho}{\beta}$ for the instanton-scale-parameter integration. Cutting off at $\rho\sim r_0=1.29\beta_0$ suggests that there are (mild) deviations from a harmonic dependence. Figure taken from \cite{HerbstHofmann2004}.\label{figsat}}
\end{figure}

The mass $m_m$ of a BPS monopole is given as \cite{BPS}
\eqb
\label{massBPSmon}
m_m=\frac{4\pi}{e}H_\infty\,,
\eqe
where $e$ denotes the defining gauge coupling of the adjoint Higgs model and $H_\infty$ the spatially asymptotic modulus of the Higgs field. Assuming that the monopole was liberated by the dissociation of a maximum-holonomy caloron at $T_0=1.32\,T_c$, the temperature where the thermodynamical (one-loop) pressure vanishes, we have \cite{HofmannBook,Diakonov2004}
\eqb
\label{eandH}
e(T_0)=12.96\,,\ \ \ \ \ H_\infty(T_0)=\pi T_0\,.
\eqe 
The electron's rest mass $m_0$ is, on one hand, given by the circular frequency $\omega_0$ of monopole-core 
oscillation, found to be equal to the mass of the two off-Cartan modes in \cite{FodorRacz2004,Forgacs}
\eqb
\label{coreoscfreq}
\omega_0=eH_\infty\,.
\eqe
On the other hand, $m_0$ decomposes into $m_m$ and the energy contained within an approximately ball-like region of radius $r_0$ (the contribution of the two vortex loops in negligible \cite{HofmannBook}, $r_0$ is denoted as $R_0$ in \cite{Hofmann2017}) due to the deconfining plasma of energy density $\rho(T_0)$
\eab
\label{massfromm0}
m_0&=&12.96\,H_\infty(T_0)=m_m+\frac{4\pi}{3}r_0^3\rho(T_0)\nonumber\\
&=&H_\infty(T_0)\left(\frac{4\pi}{12.96}+8.31\times\frac{128\pi}{3}\left(\frac{r_0}{18.31}\right)^3H_\infty^3(T_0)\right)\,.
\eae
For later use, we introduce the dimensionless temperature 
$\lambda\equiv \frac{2\pi T}{\Lambda}$ where $\Lambda$ denotes 
the Yang-Mills scale, related to $m_0$ as \cite{Hofmann2017}
\eqb
\label{YMm0}
\Lambda=\frac{1}{118.6}\,m_0\,. 
\eqe
Solving Eq.\,(\ref{massfromm0}) for $r_0$ yields
\eqb
\label{R_0solv}
r_0=4.043\,H_\infty^{-1}
\eqe
instead of $r_0=4.10\,H_\infty^{-1}$ as obtained in \cite{Hofmann2017}. It is instructive 
to compute the relative contribution of $m_m$ to $m_0$ and the ratio of Compton wave length 
$l_C=m_0^{-1}=\frac{1}{12.96\,H_\infty(T_0)}$ \cite{FodorRacz2004,Forgacs} to $r_0$ 
\eqb
\label{relm_mtom_0}
\frac{m_m}{m_0}=\frac{4\pi}{(12.96)^2}=0.0748\,,\ \ \ \ 
\frac{l_C}{r_0}=\frac{1}{52.40}\,.
\eqe 
Moreover, radius $r_0$ compares to the spatial 
coarse-graining scale $|\phi|^{-1}(T_0)$, turning the Euclidean 
time dependence of the field strength correlation within a caloron or anticaloron 
center into a mere choice of gauge \cite{HofmannBook}, as follows
\eqb
\label{R0cgsVS}
\frac{r_0}{|\phi|^{-1}(T_0)}=\frac{4.043}{\sqrt{2\left(\frac{118.6}{12.96}\right)^3}}=0.1033\,.
\eqe   
Due to a simple calculational error in Eq.\,(22) of \cite{Hofmann2017} this result is qualitatively different and implies that the ``thermodynamics'' we discussed so far actually 
occurs deep within the center of a caloron or anticaloron of a 
scale parameter $\rho\sim |\phi|^{-1}(T_0)$ (the coarse-graining 
scale of infinite-volume thermodynamics in the deconfining phase \cite{HofmannBook}). 
Since 
\eqb
\label{R0vsbeta}
\frac{r_0}{\beta_0}=1.29\ \ \ \ \ \ \ (\beta_0\equiv\frac{1}{T_0})
\eqe
Fig.\,\ref{figsat} suggests that the scale parameter integral, defining the phase of 
$\phi$ \cite{HofmannBook}, does not quite saturate a harmonic dependence on Euclidean 
time when cut off at $\rho\sim r_0$. Therefore, our above 
discussion, which assumes that such a saturation occurs within the volume $\frac{4\pi}{3} r_0^3$, is only an approximate account of the quantum physics within the self-intersection region: the monopole is always close 
to the locus of action at the inmost point of the caloron or anticaloron, rendering this 
region a highly jittery object, not only concerning the spherical shell of preconfining/confining phase that  represents the boundary of the ball but also its deep bulk.     
Still, the thermodynamical approximation employed here should yield reasonable estimates of all spatial scales and of the critical temperature $T_c=13.87/(2\pi\times 118.6)\,m_0=9.49\,$keV \cite{Hofmann2017}.  
  
\section{Lowest spherically symmetric breathing mode \label{BreathingMode}}

Let us now address a situation, where thermodynamics {\sl can be} applied without 
restrictions. Namely, we would like to compute the frequency $\Omega_0$ of 
the lowest spherically symmetric breathing mode of a deconfining SU(2) Yang-Mills 
plasma ball of homogeneous energy density $\rho$ and pressure $P$, whose temperature oscillates 
about $T_0$. Surface effects, arising from the transition between the 
deconfining (bulk) and the confining (exterior of ball) phases can be neglected 
for a sufficiently large ball mass $M\equiv\frac{4}{3}\pi R_0^3\rho(R_0)$, and 
the according expression for $\Omega_0$ reads \cite{Lamb1882,Hartland2003}
\eqb
\label{Omega}
\Omega_0=\frac{\pi c_s \Lambda}{\bar{R}_0}\,,
\eqe
where the dimensionless quantities $c_s$ (longitudinal sound velocity) and $\bar{R}_0$ ($R_0$ in units of the inverse Yang-Mills scale $\Lambda^{-1}$) are defined as
\eqb
\label{cs}
c^2_s\equiv\left.\frac{\frac{d\bar{P}}{d\lambda}}{\frac{d\bar{\rho}}{d\lambda}}\right|_{\lambda=\lambda_0}
\eqe
and $\bar{R}_0\equiv R_0\Lambda$. In Eq.\,(\ref{cs}) the use of the one-loop pressure 
$P\equiv\Lambda^4\bar{P}$ and of the one-loop energy density 
$\rho\equiv\Lambda^4\bar{\rho}$ are excellent 
approximations (modified by higher-loop corrections on the 1\%-level \cite{HofmannBook}). Amusingly, an estimate of 
$\Omega_0$ by virtue of a linearisation of the force-balance equation 
$\frac{4}{3}\pi R^3\rho(R)\ddot{R}=4\pi R^2 P(R)$ about $R_0$ and in employing energy conservation, 
\eqb
\label{enercons}
\bar{R}\equiv R\Lambda=\left(\frac{3\bar{M}}{4\pi\bar{\rho}(\bar{R})}\right)^{1/3}\ \ \ \ \ \ \ 
(\bar{M}\equiv\Lambda M),
\eqe
replaces the factor of $\pi$ in Eq.\,(\ref{Omega}) by a factor of three.  
For $\bar{P}$ and $\bar{\rho}$ we have \cite{HofmannBook}
\eab
\label{barrhoP}
\bar{P}(2a,\lambda)&\equiv&-\frac{2\lambda^4}{(2\pi)^6}\left[2\tilde{P}(0)+6\tilde{P}(2a)\right]-2\lambda\,,\nonumber\\ 
\bar{\rho}(2a,\lambda)&\equiv&\frac{2\lambda^4}{(2\pi)^6}\left[2\tilde{\rho}(0)+6\tilde{\rho}(2a)\right]+2\lambda\,,\nonumber\\ 
a&=&a(\lambda)\equiv 2\pi e(\lambda)\lambda^{-3/2}\ \ \ \ \ \ \ \ (e(\lambda_0)=12.96)\,,
\eae       
where 
\eab
\label{tilderhoP}
\tilde{P}(y)&\equiv&\int_0^\infty dx\,x^2\log\left[1-\exp\left(-\sqrt{x^2+y^2}\right)\right]\,,\nonumber\\ 
\tilde{\rho}(y)&\equiv&\int_0^\infty dx\,x^2\frac{\sqrt{x^2+y^2}}{\exp\left(\sqrt{x^2+y^2}\right)-1}\,.
\eae
Taking into account implicit (via $a(\lambda)$) and explicit dependences of 
$\bar{P}$ and $\bar{\rho}$ on $\lambda$ and employing the evolution 
equation \cite{HofmannBook}
\eqb
\label{alam}
1=-\frac{24\lambda^3}{(2\pi)^6}\left(\lambda\frac{da}{d\lambda}+a\right)a D(2a)\,,
\eqe
one derives 
\eab
\label{prhoder}
\frac{d\bar{P}}{d\lambda}&=&-
\frac{\lambda^3}{(2\pi)^6}\left(16\tilde{P}(0)+48(\tilde{P}(2a)-a^2D(2a))\right)\,,\nonumber\\ 
\frac{d\bar{\rho}}{d\lambda}&=&\frac{\lambda^3}{(2\pi)^6}\left(16\tilde{\rho}(0)+48\left(\tilde{\rho}(2a)- a^2(D(2a)-F(2a))\right)\right)+2\left(1-\frac{D(2a)-F(2a)}{D(2a)}\right)\,,\nonumber\\ 
\eae
where
\eab
D(y)&\equiv&\int_0^\infty dx\,\frac{x^2}{\sqrt{x^2+y^2}}\frac{1}{\exp\left(\sqrt{x^2+y^2}\right)-1}\,,\nonumber\\ 
F(y)&\equiv&\int_0^\infty dx\,x^2\frac{\exp\left(\sqrt{x^2+y^2}\right)}{\left(\exp\left(\sqrt{x^2+y^2}\right)-1\right)^2}\,.
\eae
Substituting Eqs.(\ref{prhoder}) into Eq.\,(\ref{cs}) at $\lambda_0=18.31$, we numerically obtain 
\eqb
\label{csnum}
c_s(\lambda_0)=0.479\,.
\eqe
For Eq.\,(\ref{Omega}) this yields 
\eqb
\label{om0num}
\Omega_0=1.506\,\frac{\Lambda}{\bar{R}_0}\,.
\eqe
Let us now compare the monopole-core induced frequency $\omega_0$ of the self-intersection 
region of the figure-eight shaped center-vortex loop (model of the electron) 
with $\Omega_0$ at one and the same radius 
\eqb
\label{radiuseq}
r_0=R_0=4.043\,H_\infty^{-1}(T_0)\,,
\eqe
see Eq.\,(\ref{R_0solv}). For $\Omega_0$ this yields
\eqb
\label{omegar0}
\Omega_0=0.372\,H_\infty(T_0)
\eqe
such that $\frac{\omega_0}{\Omega_0}=\frac{12.96}{0.372}=34.84$. This large ratio is suggestive since the 
oscillation in the self-intersection region -- quantum initiated by caloron or anticaloron action -- is induced by the classical dynamics of a monopole core \cite{FodorRacz2004,Forgacs} whose size matches the 
Compton wave length $l_C$ while the lowest symmetric breathing mode of the neutral 
deconfining ball is a consequence of sound propagation in bulk 
thermodynamics, spatially supported by a much larger system of Bohr-radius dimensions. 

Eq.\,(\ref{om0num}) is the more reliable the larger $\bar{R}_0$ is. Isotropy breaking effects, which associate with the neglected surface dynamics of the ball and/or the excitation of spherically non-symmetric oscillation states, cause this surface to (electromagnetically) radiate with a spectrum that is dominated by 
frequencies around $\nu_0\sim \frac{\Omega_0}{2\pi}$, corresponding to a wave length $l_0=\frac{1}{\nu_0}\sim \frac{2\pi R_0}{1.506}$.

\section{Summary and Discussion \label{Disc}}

This paper's purpose was to compare two situations in which a ball-like 
region of deconfining phase in SU(2) Yang-Mills thermodynamics 
oscillates about the zero of the pressure at temperature $T_0$: 
the self-intersection region of a figure-eight shaped, solitonic 
center-vortex loop (a model of the electron) containing a frequently perturbed BPS 
monopole, whose classical core dynamics drives this oscillation of (circular) frequency $\omega_0$ 
(up to a factor $\hbar$ coincident with the rest energy $m_0c^2$ of the soliton \cite{deBroglieDoc,DeBroglieRev}). Furthermore, we have considered a homogeneous, electrically neutral region whose lowest spherically symmetric oscillatory 
excitation of (circular) frequency $\Omega_0$ is supported by a finite speed of sound $c_s$, in turn based on 
the thermal ground state's quantum excitations which are microscopically mediated by 
the unit of action $\hbar$ localised within the center of a 
caloron/anticaloron \cite{HofmannBook}. At the same 
radius, $r_0=R_0=\frac{4.043}{\pi T_0}$, we obtain $\frac{\omega_0}{\Omega_0}=34.84$ 
which makes explicit the very different cause of oscillation in either case.  

We have noticed a numerical error in \cite{Hofmann2017} concerning an estimate of the system 
size $r_0$ in terms of the spatial coarse-graining scale $|\phi|^{-1}$. The correct result 
states that finite-size effects cannot be excluded at face value since the region of 
self-intersection actually is contained deeply within the ball of spatial coarse-graining corresponding to the infinite-volume situation. However, the asymptotic harmonic time dependence of the integrated field strength correlation \cite{HofmannBook}, required for the introduction of the field $\phi$, is approached to some extent when cutting the instanton-scale-parameter integration 
off at $r_0$ such that, approximately, one can still rely on 
infinite-volume thermodynamics.   

The present work only represents a first step in studying the plasma 
dynamics of a ball-like region of deconfining phase at $T_0$. More 
realistically, the physics of a certain boundary shell 
should be taken into account. The preconfining-phase part of this 
shell is superconducting (condensate of electrically charged, 
massless monopoles) \cite{HofmannBook} with implications for the 
stabilisation of the highly turbulent gas of self-intersecting vortex 
loops (electrically charged particles) within a region 
externally adjacent to this shell. Also, we did not 
address the evaporation physics (emission of non-intersecting and self-intersecting 
center-vortex loops from the surface of this shell) in case of macroscopically 
sized balls, see \cite{GH2009}, and how this process 
affects the oscillation dynamics and electromagnetic 
emission. Our results on the spherically symmetric oscillations 
of the homogeneous and macroscopic plasma could be relevant for models 
of Cepheid variables beyond the usual kappa mechanism of star dynamics and the 
description of certain, quasi-stabilised, compact and radiating objects created 
within atmospheric discharges. It is not unlikely to provide for a conceptual impact 
on plasma stabilisation in terrestial fusion experiments. \vspace{0.3cm}\\

\noindent\textbf{Acknowledgments:} We would like to acknowledge a useful 
conversation with Anton Plech.









\end{document}